\documentclass[showpacs,aps,superscriptaddress,twocolumn,amsmath]{revtex4}
\usepackage{amssymb}
\usepackage{graphicx}% Include figure files
\usepackage[percent]{overpic}
\usepackage{dcolumn}% Align table columns on decimal point
\usepackage{bm}% bold math

\begin{document}

%\preprint{APS/123-QED}
%+++++++++++++++++++++++++++++++++++++++++++++++++++++++++++
\title {Two-Dimensional PN Monolayer Sheets with Fantastic Structures and Properties}

\author{ShuangYing Ma}
\affiliation{School of Physics and Optoelectronics, Xiangtan
University, Xiangtan 411105, China}
\author{Chaoyu He}
\affiliation{School of Physics and Optoelectronics, Xiangtan
University, Xiangtan 411105, China}
\author{L. Z. Sun}
\affiliation{Hunan Provincial Key laboratory of Thin Film Materials and
Devices, School of Material Sciences and Engineering, Xiangtan
University, Xiangtan 411105, China}\email{lzsun@xtu.edu.cn}
\author{Haiping Lin}
\affiliation{Institute of Functional Nano $\&$ Soft Materials
(FUNSOM), Soochow University, Suzhou 215123, China}
\author{Youyong Li}
\affiliation{Institute of Functional Nano $\&$ Soft Materials
(FUNSOM), Soochow University, Suzhou 215123, China}
\author{K. W. Zhang}
\affiliation{School of Physics and Optoelectronics, Xiangtan
University, Xiangtan 411105, China}\email{kwzhang@xtu.edu.cn}

%++++++++++++++++++++++++++++++++++++++++++++++++++++

\begin{abstract}
\indent Three two-dimensional phosphorus nitride (PN) monolayer sheets (named as $\alpha$-, $\beta$-, and $\gamma$-PN, respectively) with fantastic structures and properties are predicted based on first-principles calculations. The $\alpha$-PN and $\gamma$-PN are buckled structure, whereas $\beta$-PN shows puckered characteristics. Their unique structures endows these atomic PN sheets with high dynamic stabilities and anisotropic mechanical properties. They are all indirect semiconductors and their band gap sensitively depends on the in-plane strain. Moreover, the nanoribbons patterned from these three PN monolayers demonstrate remarkable quantum size effect. Particularly, the Zigzag $\alpha$-PN nanoribbon shows size-dependent ferromagnetism. Their significant properties show potential in nano-electronics. The synthesis of the three phases of PN monolayer sheets is proposed theoretically, which is deserved to further study in experiments.\\
\end{abstract}\maketitle
\section{Introduction}
\indent The synthesis of Graphene and its excellent
properties\cite{1} promote the research of low-dimensional
nano-materials into two-dimensional (2D) epoch\cite{2,3,4}. In past
decade, considerable efforts have been pursued on the discovery of
2D materials beyond graphene\cite{1} such as hexagonal BN\cite{1,5},
dichalcogenide\cite{1,6}, group IV\cite{7,8,9,10}, II-VI
\cite{11,12,13}, and III-V compounds\cite{14,15,16} metastable
monolayer. Recently, as representatives of group V, the monolayer
composed purely of single element of phosphorene \cite{17,18,19,20},
arsenene\cite{21,22},
 and antimonene\cite{21}, were theoretically predicted and some of them were
 synthesized in experiment. Importantly, few-layer black phosphorus and arsenene
 show high carrier mobility\cite{23,24}, which endows them with great potential in
  future nanoelectronics. However, up to date, as the representative of group V-V
  monolayer compounds, the monolayer counterpart of phosphorus nitride compounds has
   not been reported yet even though some three-dimensional (3D) phosphorus nitride
   compounds crystals had been found several decades ago.\\
%+++++++++++++++++++++++++++++++++++++++++++++++++++++++++++
\indent To our knowledge, the only synthesized phosphorus nitride
crystal is P$_{3}$N$_{5}$ with a variety of pressure-dependent
phases\cite{25,26,27,28,29,30}. In particular, all of the phases are
only stable under high pressure except for its $\alpha$ phase. A
previous DFT calculation indicated that the $\alpha$ phase
transforms into $\gamma$ phase under around 6 GPa, and further into
$\delta$ phase with Kyanite-like structure under about 43
GPa\cite{30}. Recently, two new phases PN$_{3}$ and PN$_{2}$ were
theoretically predicted. They are also only stable under high
pressures\cite{31}. The question is why so many PN phases can only
exist under high pressures except for $\alpha$-P$_{3}$N$_{5}$.
Although Raza et al.\cite{31} pointed out that the instability of
PN$_{3}$ results from the existing of N-N units which tends to form
N$_{2}$ pair, there is not N-N bond in other PN phases. Through
structural analysis, we find that the coordination number of P and N
atom in these PN phases is approximately proportional to the stable
pressure of corresponding PN phase. For example,
$\alpha$-P$_{3}$N$_{5}$ is stable at ambient pressure\cite{31},
whose coordination number of P and N atom $\leq$ 4 and 3,
respectively; whereas the coordination
 number of P and N atom is 6 and 4, respectively for PN$_{2}$ phase whose stable
 pressure is above 200 GPa\cite{31}. Therefore, we extrapolate that to obtain a
 kind of stable phosphorus nitride compounds crystal under lower or even ambient
 pressure the coordination number of P and N atom $\leq$ 4 and 3, respectively.
 Many works had been done on phosphorus nitride compounds molecules such as cyclic
  phosphazenes with 1:1 mole fraction of P and N\cite{25,32}. The results indicated
  that cyclic phosphazenes are surprisingly stable, and their stability is insensitive
  to their planar or puckering configuration\cite{32}. Meanwhile, since N and P atoms
  are favorable to form three coordinated 2D systems, such as h-BN monolayer\cite{1,5}
   and phosphorene\cite{17,18,19,20}, adopting hexagonal three coordinated PN units
   with 1:1 mole fraction between P and N is hoping to construct 2D PN monolayer nano
    materials.\\
%++++++++++++++++++++++++++++++++++++++++++++++++++++++++++++
\indent In this paper, using first-principles method we propose three stable 2D monolayer
 phases of phosphorus nitride named as $\alpha$-, $\beta$-, and $\gamma$-PN. They are all
 indirect band gap semiconductor with low-buckled honeycomb structures. Their band gaps
 can be effectively tailored by in-layer strain, patterning, and multi-layer stacking.
 Meanwhile, we propose PN compound closely similar to graphite based on the 2D monolayer
  PN obtained in our present work due to the weak inter-layer interaction for the three
  monolayer PN.\\
%#######################################################################
\section{Computational method}
\indent We performed first-principles calculations based on density functional theory as implemented in the VASP code\cite{33,34} to investigate the equilibrium structures, stability, and electronic properties of $\alpha$-, $\beta$-, and $\gamma$-PN. The electron-electron interaction was treated with a generalized gradient approximation (GGA) proposed by Perdew, Burke, and Ernzerhof (PBE)\cite{35}. Projector-augmented wave (PAW) \cite{36,37} method was used for describing interaction between valence electrons and core. A kinetic-energy cutoff of 400 eV was selected for the plane wave basis set. To avoid the interaction between neighboring images a vacuum space of 15 {\AA} perpendicular to the plane of the 2D systems was set. The Brillouin zone was sampled using 13$\times$13$\times$1 Monkhorst-Pack k-point scheme. The total energy convergence criterium was $10^{-5}$ eV. All systems were fully relaxed until the residual Hellmann-Feynman forces were smaller than 0.01 eV/{\AA}. The energetic stability of the systems was evaluated by comparing their formation enthalpy\cite{38,39} $\Delta G$ defined as,
%++++++++++++++++++++++++++++++++++++++++++++++++++
\begin{eqnarray}\label{equ1}
\Delta G=E_{tot}-\sum\chi_{i}\mu_{i}.
\end{eqnarray}
%++++++++++++++++++++++++++++++++++++++++++++++++++
The term $E_{tot}$ is the cohesive energy per atom of the specific PN phase considered here, $\chi_{i}$ (i denote P, N) is the molar fraction of the groups and they obey the rule of $\sum\chi_{i}=1$. $\mu_{i}$ is the chemical potential of the constituents at a given state. We chose $\mu_{P}$ as the cohesive energy per atom of pristine monolayer black phosphorene; and $\mu_{N}$ was taken as the binding energy per atom of $N_{2}$ molecule. To verify the stabilities of the three phases, besides phonon spectrum, ab initio molecular dynamics (AIMD) simulations were also performed.\\
%+++++++++++++++++++++++++++++++++++++++++++++++++++++++++++++++++++++++++
\section{Results and Discussion}
%+++++++++++++++++++++++++++++++++++++++++++++++++++++++++++++++++++++++++
\indent With the restriction as mentioned above:(i) 1:1 mole fraction between N and P; (ii) two dimensional hexagonal configuration, we propose three two-dimensional phosphorus nitride (PN) monolayer sheets named as $\alpha$-, $\beta$-, and $\gamma$-PN, respectively. The optimized structures of the three phases are shown in Fig. 1. The structural parameters of the three PN sheets are listed in Tab. 1. Due to the different local environment of P and N atoms, there are two and three non-equivalent bond lengths for $\alpha$ and $\beta$ phases, respectively. The bond length between P and N atom of the three phases is around 1.725$\sim$1.821 \r{A}. There are respectively four and eight atoms in the rectangular unit cell of $\alpha$- and $\beta$-PN, and two atoms in hexagonal unit cell of $\gamma$-PN. Each atom in the three phases shows threefold coordination. All of them show non-planar configuration with a thickness about 0.861$\sim$1.897 \r{A} due to the phosphorus pro to form a tetrahedral configuration with its three nitrogen neighbors. However, as for $\alpha$- and $\beta$-PN each nitrogen and its three nearest neighbor (NN) phosphorus are nearly on an identical plane. The formation enthalpy as listed in Tab. 1 shows that $\alpha$ and $\beta$ phases are more stable than that of $\gamma$ phase. The results indicate that as for PN monolayer phase the P and N atoms are energetically favorable to form $sp^3$ and $sp^2$ hybridized configuration, respectively. The structural difference of the three monolayer PN sheets can be ascribed to the different way of connecting tetrahedral coordinated phosphorus atom with its three NN nitrogen atoms. The side view of the $\beta$-PN shows ridge-like configuration that is, to some extent, analogous to that of $\alpha$- and $\delta$-phosphorene. $\alpha$- and $\gamma$-PN are similar to the configuration of $\beta$- and $\gamma$-phosphorene, respectively.\\

%+++++++++++++++++++++++++++++++++++++++++++++++++++++++++++++++++++++++++
\begin{figure}
\includegraphics[width=3.1in]{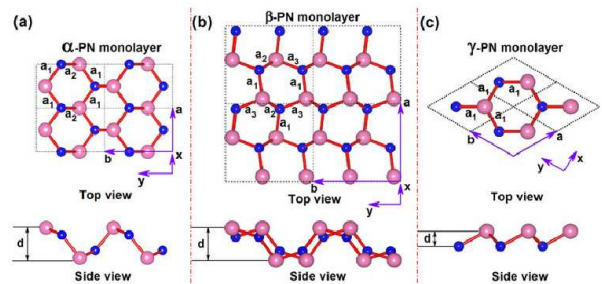}\\
 \caption{(a), (b), and (c) respectively are top and side view of 2$\times$2 supercell of optimized structures of $\alpha$-, $\beta$-, and $\gamma$-PN monolayers. The balls in blue and pink represent nitrogen and phosphorus atoms, respectively. $d$ is the thickness of the quasi-2D monolayers. a$_1$, a$_2$, and a$_3$ stand for nonequivalent bond of the three phases.} \label{fig1}
\end{figure}
%+++++++++++++++++++++++++++++++++++++++++++++++++++++++++++++++++++++++++
\begin{table*}
\caption{The lattice constants (a and b), nonequivalent bond length between phosphorus and nitrogen atom (a$_1$, a$_2$ and a$_3$), thickness of the monolayers \emph{d}, space group (SG), formation enthalpy ($\Delta$G defined by equation 1) of $\alpha$-, $\beta$-, and $\gamma$-PN monolayers.}\label{tab1}
\begin{tabular}{cccccccccc}
\hline\hline
 & $\vec{a}$(\AA) & $\vec{b}$(\AA) & a$_1$(\AA) & a$_2$(\AA) & a$_3$(\AA)& SG & \emph{d}(\AA)&$\Delta$G(eV)\\
\hline
$\alpha$-PN    &2.703  &4.190  &1.725  &1.821  &       &PMN21     &1.892 &-0.40\\
$\beta$-PN     &4.516  &5.070  &1.749  &1.797  &1.737  &PCA21     &1.897 &-0.36\\
$\gamma$-PN    &2.756  &2.756  &1.809  &1.809  &       &P3M1      &0.861 &-0.08\\
\hline\hline
\end{tabular}
\end{table*}
%+++++++++++++++++++++++++++++++++++++++++++++++++++++++++++++++++++++++++
%+++++++++++++++++++++++++++++++++++++++++++++++++++++++++++++++++++++++++
\indent The formation enthalpy ($\Delta$G) of the three PN phases as shown in Tab. 1
is negative suggesting their possible existence in reality. The most stable phase is
 $\alpha$-PN which is 40 meV more stable than that of $\beta$-PN. To further confirm
 their dynamic stability, we calculated their phonon band structures and phonon density
 of states\cite{40}. The phonon spectra are shown in Fig. 2. There are no negative
 frequencies and states in the phonon band structure and phonon density of states,
 confirming the dynamic stability of the three PN sheets. The highest frequency of
 $\alpha$-, $\beta$-, and $\gamma$-PN, as an indication of the robustness of the P-N
 bonds, reaches up to 26.8, 26.0, and 22.3 THz, respectively, which to some extent
 indicates their relative stability. The higher frequency of longitudinal optical modes
  manifests a larger in-plane rigidity. The results indicate that the rigidity sequence
   is $\alpha$ $>$ $\beta$ $>$ $\gamma$. To further check their thermodynamic stabilities,
    ab initio molecular dynamics (AIMD) calculations were performed. In the calculations,
     relatively large supercells (5$\times$4$\times$1, 1$\times$3$\times$3,
     and 4$\times$4$\times$1 for $\alpha$-, $\beta$-, and $\gamma$-PN, respectively),
     NVT ensemble, and 1 fs time step were adopted. The results indicate that the three
      PN sheets can sustain their original structures at least 6 ps under 800 K. In
      particular, the $\alpha$-PN even does not show structural instability during a 6
      ps AIMD simulation at 1600 K. The above results give firm evidence that the three
       PN sheets are stable enough to be observed in the experiments.\\
%+++++++++++++++++++++++++++++++++++++++++++++++++++++++++++++++++++++
\indent We then investigate the slopes of the longitudinal acoustic branches near $\Gamma$ point, which corresponds to the speed of sound and in-plane stiffness. According to the results in Fig. 2 (a), the sound velocity along the $\Gamma$-$X$ orientation of $\alpha$-PN ($\nu_s^{\Gamma-X}$=11.2km/s) is nearly two times larger than that along $\Gamma$-$Y$ direction ($\nu _s^{\Gamma-Y}$=5.3 km/s). The results indicate that for $\alpha$-PN, the stiffness is anisotropic that along $\vec{a}$ orientation is much larger than that along $\vec{b}$ direction. If a finite compression stress is applied along $\vec{a}$ direction the bond bending is prevail to bond stretching due to the lower energy cost for the former one. Similar results can be found for $\beta$-PN. The velocity of sound along the $\Gamma$-$Y$ direction of $\beta$-PN ($\nu_s^{\Gamma-Y}$=6.0 km/s) is nearly the half of that along $\Gamma$-$X$ orientation ($\nu _s^{\Gamma-X}$=10.54 km/s). Accordingly, the stiffness along $\vec{b}$ orientation is much lower than that along $\vec{a}$ direction. Considering the structure equivalence of $\gamma$-PN along $\vec{a}$ and $\vec{b}$ axis of unit cell, we only present the velocity of sound along one crystal axis direction. Corresponding to the reciprocal space, the direction is along $\Gamma$-$M$. The results indicate that the $\nu_s^{\Gamma-M}$ is 6.89 km/s. Meanwhile, the velocity of sound along the diagonal direction of crystal axis $\vec{a}$ and $\vec{b}$, namely, $\nu_s^{\Gamma-K}$=7.06 km/s, which is closed to that along crystal axis direction. The results derive from the isotropic structure characteristics of $\gamma$-PN.\\
%+++++++++++++++++++++++++++++++++++++++++++++++++++++++++++++++++++++++++
\begin{figure*}[htbp]
 \includegraphics[width=6.5in]{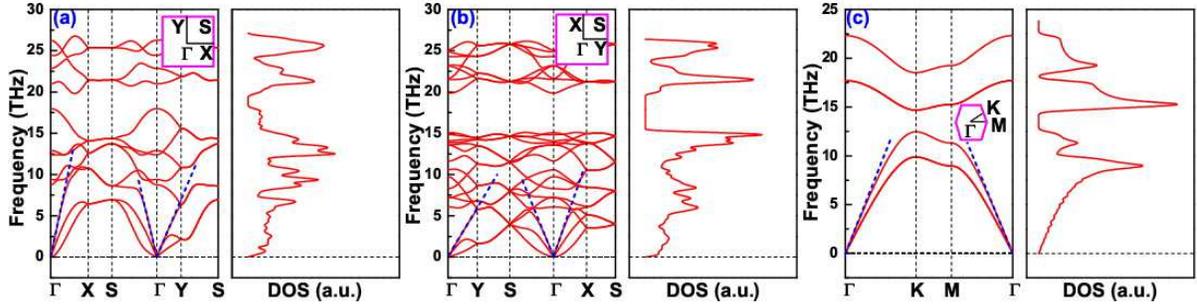}\\
 \caption{Phonon band structures and density of states of (a) $\alpha$-PN,(b) $\beta$-PN, and (c) $\gamma$-PN, respectively.}
 \label{fig2}
 \end{figure*}
%+++++++++++++++++++++++++++++++++++++++++++++++++++++++++++++++++++++++++
%+++++++++++++++++++++++++++++++++++++++++++++++++++++++++++++++++++++++++
\begin{figure}[htbp]
 \includegraphics[width=3.2in]{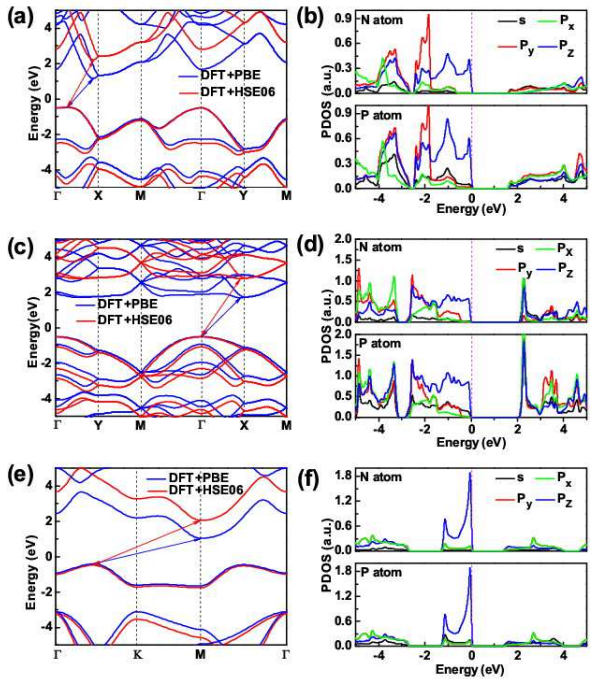}\\
 \caption{(a), (c), and (e) respectively are the band structures of $\alpha$-PN, $\beta$-PN, and $\gamma$-PN. (b), (d), and (f) correspond to partial density of states of $\alpha$-, $\beta$-, and $\gamma$-PN, respectively.}
 \label{fig3}
 \end{figure}
%+++++++++++++++++++++++++++++++++++++++++++++++++++++++++++++++++++++++++
\indent As shown in Fig. 3 (a), (c), and (e), the DFT-PBE calculations show that the
single-layer $\alpha$-PN, $\beta$-PN, and $\gamma$-PN are all indirect band gap
semiconductor with the energy gaps of 1.638, 2.197, and 1.800 eV, respectively.
Considering the fact that PBE functional always underestimates the band gap of
semiconducting materials, we also use HSE06 functional\cite{41,42} to study their
energy gaps. The revised energy-gap values are 2.689, 3.282, and 2.532 eV, respectively.
Partial density of states (PDOS) for the three phases are shown in
Fig. 3 (b), (d), and (f). The results indicate that the valence-band
maximum (VBM) of $\alpha$-PN is mainly contributed by $\emph{p}_z$ states of N and P atoms
and s orbital of P atoms; for $\beta$-PN, the VBM primarily derives from $\emph{p}_z$ orbital
 of N and P atoms and s orbital of P atoms; the VBM of $\gamma$-PN is largely determined
 by $\emph{p}_z$+$\emph{p}_x$ orbitals of N and P and s orbital of P atoms. Below VBM,
 the $\emph{p}$ and s orbitals of N and P atoms show similar electronic resonance,
 suggesting strong hybridization and charge transfer between N and P atom. The charge
  transfer between N and P atom can be clearly verified by the charge density difference
  as shown in Fig. 4 (a), (b), and (c). The results show that electrons transfer from P
  to N atom and mushroom-shaped electron cloud is generated for P atom due to charge
  redistribution. Bader charge analysis\cite{43,44} reveals the transferred net charge
   for $\alpha$-, $\beta$-, and $\gamma$-PN is 3.257, 3.221 and 3.110e, respectively,
   indicating obvious ionic bond characteristics between P and N atom.\\
%++++++++++++++++++++++++++++++++++++++++++++++++
\begin{figure}[htbp]
 \includegraphics[width=3.3in]{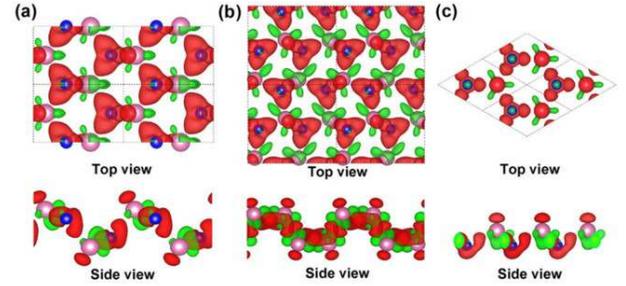}\\
 \caption{Charge density difference of $\alpha$-PN (a), $\beta$-PN (b), and
  $\gamma$-PN (c), respectively. The isosurface is 0.018 e/{\AA}$^3$). The red and green color stands for charge increase and decrease region corresponding to isolated atoms, respectively.}
 \label{fig4}
 \end{figure}
%+++++++++++++++++++++++++++++++++++++++++++++++++++++++++++++++++++++++++
\begin{figure}[htbp]
 \includegraphics[width=3.3in]{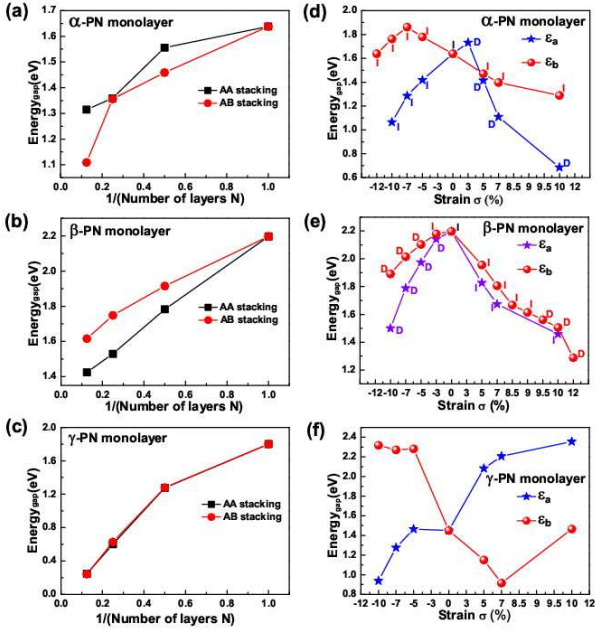}\\
 \caption{(a), (b), and (c) respectively are the band gap of $\alpha$-, $\beta$-, and $\gamma$-PN in function of the stacking thickness. (d), (e), and (f) shows the band gap in function of strain for $\alpha$-, $\beta$-, and $\gamma$-PN, respectively. I and D denote the indirect and direct band gap semiconductor, respectively.}
 \label{fig5}
 \end{figure}
%+++++++++++++++++++++++++++++++++++++++++++++++++++++++++++++++++++++++++
\begin{figure}[htbp]
 \includegraphics[width=3.0in]{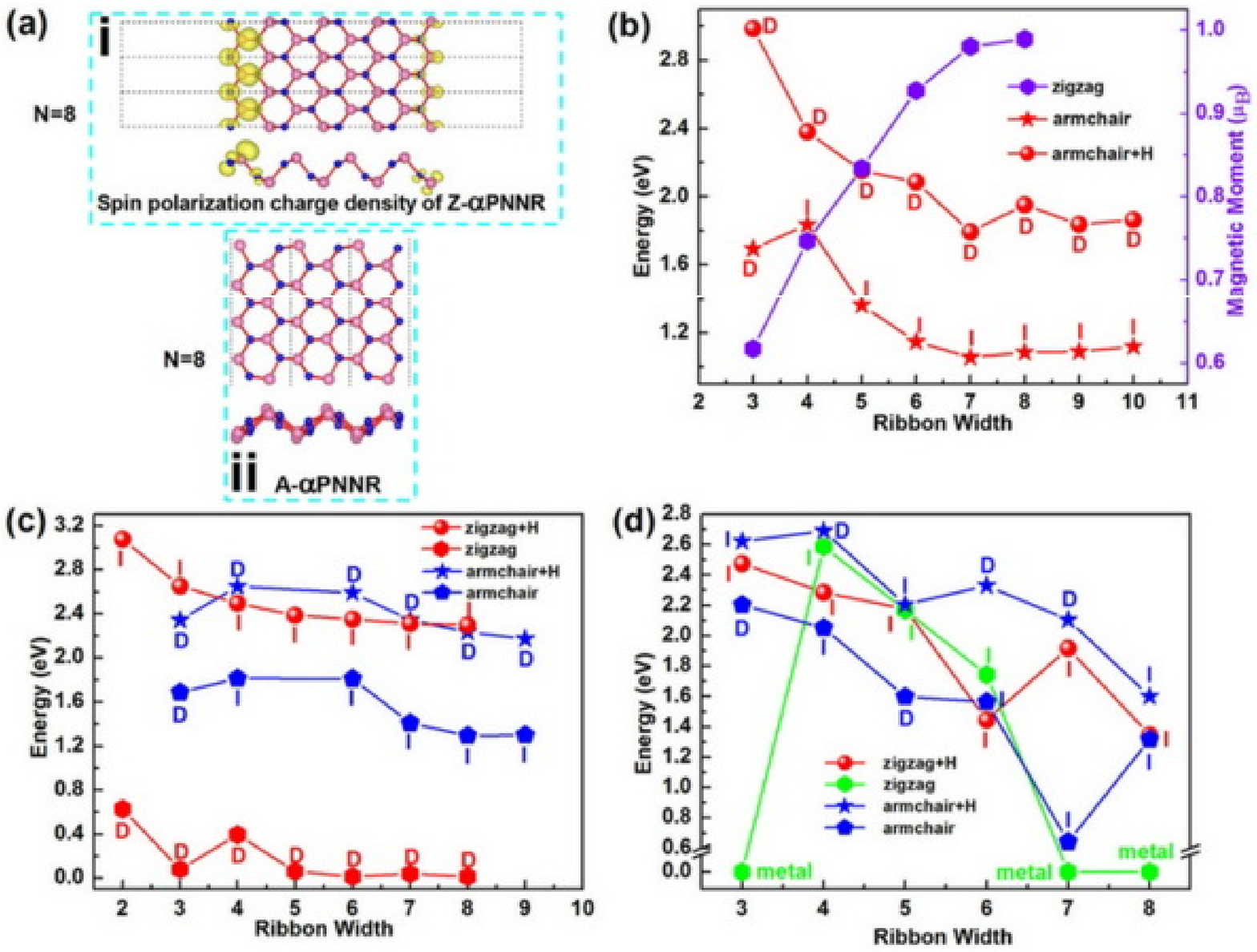}\\
\caption{Panel i and ii in (a) are the DFT-optimized structures of
Z-$\alpha$PNNR and A-$\alpha$PNNR with the width of N=8,
respectively. The yellow color of panel i is the spin polarized
charge density with isosurface of 0.005e/{\AA}$^3$. (b), (c), and
(d) are the band gap of nanoribbon as a function of ribbon width (N)
for $\alpha$-PN, $\beta$-PN, and $\gamma$-PN, respectively. The
letter I and D in the figures denote the indirect and direct
semiconductor, respectively. The symbol ``+H" means the edges of the
 nanoribbon is passivated by H atoms.}
 \label{fig6}
 \end{figure}
%+++++++++++++++++++++++++++++++++++++++++++++++++++++++++++++++++++++++++
\indent Based on the three 2D monolayer PN sheets proposed above, we investigate
the 3D PN systems analogous to graphite through stacking of them. Fig. S1 shows the
 optimized geometries with different stacking orders for the 3 kinds of multilayer
 PN phases. The minimum interlayer distance for multilayer PN phases is about 4.343 {\AA},
 which is at least 0.848 {\AA} larger than that of AA stacking black Phosphorene. Moreover,
  except for a small minish of the lattice constants and bond length for multilayer
  $\gamma$-PN, the structure parameters of the other two multilayer PN phases are nearly
   unchanged. The band gap of the three multilayer phases as shown in Fig. 5 (a), (b),
   and (c) is approximately inversely proportional to the number N of stacking layer.
   Meanwhile, the dependence of the band gap on the stacking layer is sensitive to the
   stacking order except for the $\gamma$ phase. Although the van der Waals interaction
    between layers will result in the energy level splitting and the strength of the
    splitting is proportional to N, the band-structure of multilayer PN are roughly
    similar to their corresponding monolayer PN phases, as shown in Fig. S2. In contrast
     to the case of phosphorene whose energy gap decreases with the increase in N is
     the reason of wave function overlap between layers, the wave function overlap of
      multilayer PN is slight (the results are shown in Fig. S3). The results indicate
      that the dependence of the band gap on the N of multilayer PN derives from van der
       Waals interactions between layers due to its larger interlayer distance.
       For $\alpha$-PN, the band gap can be tuned from 1.64 eV (monolayer) to 1.31
       (AA stacking, N=8) or 1.11 eV (AB stacking, N=8). As for $\beta$-PN, its band
        gap changes from 2.20 eV (monolayer) to 1.42 eV (AA stacking, N=8) or 1.62
         eV (AB stacking, N=8). The band gap of $\gamma$-PN for AA and AB stacking
         is nearly degenerate as shown in Fig. S2 due to its larger interlayer distance.
          The tunable band gap ranges from 1.80 eV (monolayer) to 0.24 eV (N=8). The
           dependence of the band gap on the thickness produces the three PN phases
           show wide range adjustability of their electronic properties.\\
%+++++++++++++++++++++++++++++++++++++++++++++++++++++++++++++++++++++++++
\indent Another intriguing issue is the sensitive dependence of the band gap of the three PN phases on in-plane strain exerted along two axial directions, as shown in Fig. 5 (d), (e), and (f) for $\alpha$-PN, $\beta$-PN, and $\gamma$-PN, respectively. Considering the un-planar feature of the three phases, the external strain -12\%$\sim$12\% can be achieved without large energy cost similar to that of phosphorene\cite{17,19}. $\alpha$-PN is an indirect-band-gap semiconductor under zero strain. When apply strain along axial vector $\vec{a}$, its energy gap decreases regardless of compression or tension. Meanwhile, $\alpha$-PN transforms into a direct band gap semiconductor when the stretching strain is larger than 3\%. Such band type change is suitable for the application of $\alpha$-PN in optoelectronics. Considering the strain scope, it can be easily exerted on the system when $\alpha$-PN is grown on substrate. The band structures of $\alpha$-PN as shown in Fig. S4 (a) indicate that its VBM and CBM turn from original $\Gamma$-X line to $\Gamma$ point as increasing of tensile strain along the a axis. When the $\sigma$ is larger than 3\% its band gap turns from indirect to direct one. The energy gap is also sensitive to the strain which is tuned from 1.638 eV at $\sigma=0$ to 1.065 eV at $\sigma=-10\%$ and 0.686 eV at $\sigma=10\%$. However, exerting stress along another lattice vector $\vec{b}$, whether compression or stretch, the band of $\alpha$-PN remains its indirect type, as shown in Fig. S4 (b). The energy gap increases from 1.638 eV under $\sigma=-12\%$ to 1.861 eV under $\sigma=-7\%$, whereas the band gap decreases from 1.861 eV at $\sigma=-7\%$ to 1.288 eV at $\sigma=10\%$. As for $\beta$-PN, its band gap decreases for both compressed and stretched strain along both cell vectors. By compressing along vector $\vec{b}$ or $\vec{a}$, the gap can be modified from 2.197 eV (at $\sigma=0$) to 1.891 eV (along $\vec{b}$, for $\sigma=-10\%$) and 1.288 eV (along $\vec{b}$, for $\sigma=12\%$) or 1.500 eV (along $\vec{a}$, for $\sigma=-10\%$) and 1.459 eV (along $\vec{a}$, for $\sigma=12\%$). When the strain along $\vec{a}$, the system behaves as indirect and direct band gap semiconductor for stretch and compress strain, respectively. However, as for the strain being along $\vec{b}$, when the strain is larger than 9\% and less than -5\%, the system behaves as direct band gap semiconductor. The band structures of $\beta$-PN as shown in Fig. S5 indicate that the VBM of $\beta$-PN locates at $\Gamma$ point under zero stain. The occurrence of the indirect-direct transformation is determined by the location of CBM. The CBM of $\beta$-PN under zero strain is close to X point. By applying suitable strain along a specific axis, the location of the CBM shifts from X to $\Gamma$ point producing direct band gap semiconductor. As for $\gamma$-PN, its type of band gap is un-effected by the strain along both cell vector. When the strain is along vector $\vec{a}$ within the range of (10\%, -10\%), its energy gap change from 2.356 to 0.938 eV. However, when the strain is along vector $\vec{b}$, the response of the band gap to the strain is reversed to that along vector $\vec{a}$. Briefly summarized, external strain can effectively tune the band gap and even band type of the three 2D PN sheets. Such feature of the three systems is significant for their application in nanoelectronics and optoelectronics.\\
%+++++++++++++++++++++++++++++++++++++++++++++++++++++++++++++++++++++++++
\indent Patterning one dimensional (1D) nanoribbons from 2D system, such as graphene nanoribbon (GNR), is an effective approach to modulate the electronic properties of the 2D system and promote its applications. Considering the zigzag and armchair conventional notation of GNR, we take the zigzag and armchair edge PN nanoribbon (Z-PNNR and A-PNNR) as examples to study the electronic properties of 1D PNNR. Meanwhile, the number N of dimer lines across a specific nanoribbon is adopted to denote the width of nanoribbon with zigzag and armchair edges. For example, the Z- and A-$\alpha$PNNR with the width N=8 is shown in panel i and ii of Fig. 6 (a), respectively. Both bared and H passivated edges of PNNR are considered in present work. Z-$\alpha$PNNR and A-$\alpha$PNNR can be fabricated by tailoring the layer along the $\vec{a}$ and $\vec{b}$ vector of $\alpha$-PN, respectively. Our results indicate that among all PNNRs studied in our present work only the edges of Z-$\alpha$PNNRs can not be passivated by H atoms. Such feature produces the Z-$\alpha$PNNRs behave as ferromagnetic metal. As is shown in Fig. 6 (b), the magnetic moment of Z-$\alpha$PNNRs increases linearly with the increase in the ribbon width. The magnetic moment saturates to 1 $\mu B$ when N$>$6. The spin polarization charge density as shown in panel i of Fig. 6 (a) indicates that the spin coupling between two edges is ferromagnetic which is different from that of zigzag GNR. Meanwhile, the net magnetic moment is dominantly contributed by the edge N atoms and their nearest neighbor P atoms. A-$\alpha$PNNRs are all semiconductors whether their edges are H-passivated or not, as shown in Fig. 6 (b). Moreover, their energy gap decreases with the increase in the ribbon width. Interestingly, by passivating the edges with H atoms, their band type transforms from indirect one into direct type. Meanwhile their energy gaps remarkably increase with the H passivation. In the case of $\beta$-PNNR, the situation of A-$\beta$PNNRs is very similar to that of A-$\alpha$PNNRs. However, the edge-bared Z-$\alpha$PNNRs are all direct-semiconductors. Their band type can be transformed into indirect one once the edges passivated by H atoms. The energy gap of Z-$\alpha$PNNRs with bared edges tend to equilibrate around several meV when N $>$5. However, the H passivation significantly increases its band gap to around 2.4 eV. As for the case of $\gamma$-PNNRs, the H passivation can reverse the band type of A-$\gamma$PNNRs when the band width is smaller than 8. The energy gap of all type of $\gamma$-PNNRs basically decreases with the increase in the ribbon width. When the ribbon width of bared edge Z-$\gamma$PNNRs is smaller than 4 and larger then 8, the ribbon turn to metal.\\
%+++++++++++++++++++++++++++++++++++++++++++++++++++++++++++++++++++++
\begin{figure}[htbp]
 \includegraphics[width=3.2in]{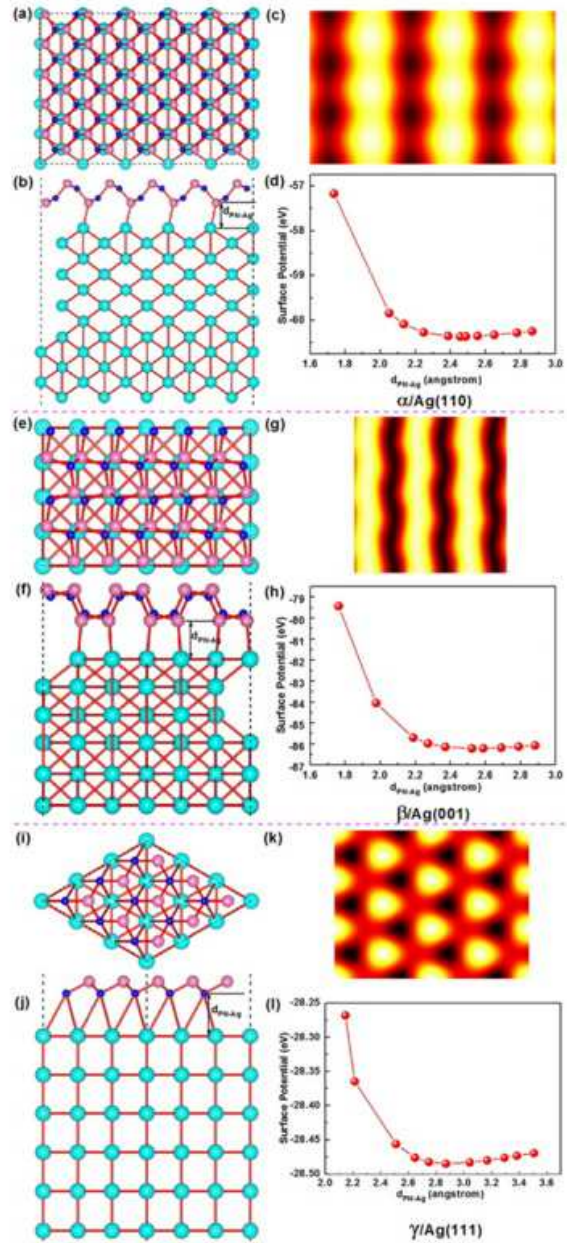}\\
 \caption{(a) and (b), (e) and (f) as well as (i) and (j) stands for the top and side view
 of $\alpha$-PN/Ag(110), $\beta$-PN/Ag(001) and $\gamma$-PN/Ag(111) system,
 respectively. (c), (g), and (k) are respectively the
 corresponding STM mapping of $\alpha$-PN/Ag(110), $\beta$-PN/Ag(001) and $\gamma$-PN/Ag(111)
 systems under -1.0 V voltage. (d), (h) and (i) shows the surface potential of
$\alpha$-PN/Ag(110), $\beta$-PN/Ag(001)
  and $\gamma$-PN/Ag(111) systems, respectively.}
 \label{fig7}
 \end{figure}
%+++++++++++++++++++++++++++++++++++++++++++++++++++++++++++++++++++++++++
\indent Similar to the synthesis of graphene\cite{45} on metal surfaces, we here propose
 a possible growth method for the three monolayer PN sheets on Ag substrate by CVD method
 with cyclic phosphazenes. Although Raza et al.\cite{31} indicated that extreme temperatures
  and pressures are required for synthesis many nitride compounds due to the high kinetic
   barrier to polymerisation derived from the high stability of nitrogen molecule at ambient
    pressure and the strong N$\equiv$N triple bond, we can overcome these difficulties
     through directly adopting some PN compounds molecules rather than N$_{2}$. The growth
      of the $\alpha$-, $\beta$-, and $\gamma$-PN monolayer on Ag(110), Ag(001), and Ag(111),
       respectively are shown in Fig. 7 (a) and (b), (e) and (f), and (i) and (j),
       respectively. The lattice mismatch between $\alpha$-PN and Ag(110) is 6.89\%
       and 2.48\% along vector $\vec{a}$ and $\vec{b}$, respectively. For
       $\gamma$-PN/Ag(111), the lattice-misfit is about 3.18\%. The relatively small
        mismatch between $\alpha$- and $\gamma$-PN and Ag substrates implies the potential
         realization of them in experiments. Although the lattice mismatch for
         $\beta$-PN/Ag(001) system is larger than 13\%, which is a little wild, we
         can still expect its realization similar to the growth of TiC monolayer on
          NiO(001) surface \cite{46}. To better guide experiments, we further calculated
          surface potential (SP) between Ag substrate and the monolayer PN. The
          energetically favorable spacing between monolayer PN and Ag substrate
          is 2.455, 2.551 and 3.307 {\AA} for $\alpha$-PN/Ag(110), $\beta$-PN/Ag(001),
          and $\gamma$-PN/Ag(111) systems, respectively. The relative large distance
          between the monolayer PN and Ag substrate indicates their weak interaction.
          The excellent properties of the monolayer PN is hoping to be reserved under
          such growth condition. Meanwhile, we simulate their STM images when the
          monolayer grows on the Ag substrate. we only presented the corresponding STM
           images of the three systems at -1.0 V voltage, as shown in in Fig. 7 (c),
           (g) and (k), respectively, due to the results at two voltages are nearly the
           same. Obviously, the STM images clearly provide the non-planar and periodic
            structural characteristics of the three monolayer PN sheets as useful
             references for experiments.\\
%+++++++++++++++++++++++++++++++++++++++++++++++++++++++++++++++++++++++++
%+++++++++++++++++++++++++++++++++++++++++++++++++++++++++++++++++++++++++
\section{Conclusions}
\indent Using first-principles calculations we predict three novel 2D phosphorus nitride (PN) monolayer sheets that can be stable under ambient pressure in contrast to previously reported 3D PN crystals. The dynamic stability of the three 2D PN even reach to high temperature under ambient pressure. The $\alpha$- and $\gamma$-PN show buckled configuration while $\beta$-PN shows puckered one. The three phases are all indirect semiconductors. Their band gap and band type can be effectively modulated by multilayer stacking, in-plane
strain, and 1D patterning. Particularly, the Z-$\alpha$PNNRs show size-dependent ferromagnetism.\\
%+++++++++++++++++++++++++++++++++++++++++++++++++++++++++++++++++++++++++
\begin{acknowledgments}
This work was financially supported by the National Natural Science
Foundation of China (Grant Nos. 11574260 and 11274262) and
the Natural Science Foundation of Hunan Province, China (Grand No.
14JJ2046).
\end{acknowledgments}
%+++++++++++++++++++++++++++++++++++++++++++++++++++++++++++++++++++++++++

\end{document}